\documentclass[11pt,a4paper]{article}
\usepackage[T1]{fontenc}
\usepackage[latin1]{inputenc}
\usepackage{amsmath,amssymb,amsthm}
\usepackage{epsfig,graphicx}
\def\be{\begin{equation}}
\def\ee{\end{equation}}
\def\bea{\begin{eqnarray}}
\def\eea{\end{eqnarray}}

\newcommand{\rme}{\mathrm{e}}
\newcommand{\rmi}{\mathrm{i}}

\begin{document}

\begin{center}
{\Large \bf Area distribution of two-dimensional random walks \\ on a square lattice}\\[0.5cm]

{\large \bf Stefan Mashkevich}\footnote{mash@mashke.org}\\[0.1cm]
Schr\"odinger, 120 West 45th St., New York, NY 10036, USA \\ and \\
Bogolyubov Insitute for Theoretical Physics, 03143 Kiev, Ukraine\\[0.4cm]
{\large \bf St\'ephane Ouvry}\footnote{ouvry@lptms.u-psud.fr}\\[0.1cm]
Universit\'e Paris-Sud, Laboratoire de Physique Th\'eorique et Mod\`eles
Statistiques\footnote{Unit\'e Mixte de Recherche CNRS-Paris Sud, UMR 8626}\\
91405 Orsay, France
\\[0.2cm]

\today

\end{center}

\vskip 0.5cm
\centerline{\large \bf Abstract}
\vskip 0.2cm
The algebraic area probability distribution of closed planar random walks
of length $N$ on a square lattice is considered.
The generating function for the distribution satisfies a
recurrence relation in which the combinatorics is encoded.
A particular case generalizes the $q$-binomial theorem to the case of three addends.
The distribution fits the L\'evy probability distribution for Brownian curves
with its first-order $1/N$ correction quite well, even for $N$ rather small.


\section{Introduction}

We are considering closed planar random walks on a square lattice.
We aim at finding the probability distribution of the algebraic area $A$
enclosed by a random walk of length $N$, starting and ending at the origin.

Random walks are a discretized version of closed continuous
Brownian curves, which are obtained in the limit $N\to\infty$. In this limit 
the average size of the walk diverges as $\sqrt N$,
its average area  is proportional to $N$,
therefore its finite renormalized area is $a = A/N$.
As first shown by L\'evy \cite{Levy},
the asymptotic probability distribution is
\be
P_{N\to \infty}(a) = \frac{\pi}{\cosh^2 (2\pi a)} \;.
\ee
Discrete random walks whose links can point in arbitrary directions
were previously considered \cite{Brereton}, and the L\'evy distribution
recovered in the continuous limit \cite{Khandekar}.
Certain analytic results were also obtained
concerning the area distribution of walks in the presence of
random traps \cite{Samokhin},
of directed random walks (those that begin and end
on the $y=0$ line) \cite{Jonsson},
as well as walks confined to a finite-size box \cite{Desbois}.

The problem of the random walks area distribution and   $n$-winding sectors  area distribution  
on a square lattice arises, for example, in the context of random magnetic
impurities and the integer quantum Hall effect ~\cite{Desboisbis}.
Also, the distribution 
is connected to the partition function of a
lattice gauge model with Z(2) gauge group
interacting with a Z(2)-valued Higgs field \cite{Borisenko}.

The problem has been attacked from two sides.
For big $N$, a finite-size correction to the L\'evy distribution
at first order in $1/N$ was derived in Ref.~\cite{Bell}
[see Eq.~(\ref{mapping}) below] by relating the
number of walks to the trace of the Hamiltonian of
the Harper model \cite{Harper},
which is in turn related to the Hofstadter model \cite{Hofstadter}
of an electron moving on a two-dimensional square lattice
in the presence of a uniform magnetic field orthogonal to the plane.
On the other hand, for a finite $N$, all the probabilities
involved are rational numbers, therefore the sought distribution
is a rational function.
Properties of its $k$-th moment --- which turns out to be
a rational function of $N$ with integer coefficients ---
have been studied in Ref.~\cite{Mingo},
where the L\'evy distribution was also explicitly obtained in the $N\to\infty$ limit (see also \cite{Beguin}).
An exact expression for the finite $N$  distribution
has, however, remained out of reach.

In this paper we derive a recurrence relation
for the generating function of the area probability distribution
for an arbitrary $N$, which is interpreted within a statistical mechanics
approach as well as in terms of $q$-commuting operators.
The asymptotic limit of the distribution and its first-order $1/N$-correction
are reproduced numerically.

\section{Generating function of the area distribution}

Denote links on the lattice pointing right, up, left, and down
with operators $x$, $y$, $x^{-1}$, $y^{-1}$, respectively.
By convention, all walks begin at the origin.
A walk of length $N$ is then defined by 
a sequence of links $\{l_1, \ldots, l_N\}$,
where each $l_k$ can be one of the four operators above.
For a walk to be closed,
the number of $x$'s has to be equal to the number of $x^{-1}$'s,
the same for the $y$'s and $y^{-1}$'s, hence an even $N$.

\sloppypar In order to calculate the algebraic area $A(l_1, \ldots, l_N)$
enclosed by a walk (positive/negative when encircled anticlockwise/clockwise), it is sufficient to
note that for two walks differing only by an interchange of a pair of
subsequent links,
$A(l_1, \ldots, l_{k-1}, x, y, l_{k+2}, \ldots, l_N)
= A(l_1, \ldots, l_{k-1}, y, x, l_{k+2}, \ldots, l_N) + 1$.
Introduce the $q$-commutator
\be
xy = qyx \;, \label{qcomm}
\ee
where $q \ne 0$ and $q \ne \pm 1$.
The full commutation table follows immediately:
e.g., multiplying both sides by $x^{-1}$ on the left and on the right
yields $yx^{-1} = qx^{-1}y$, etc.
Then the algebraic area of a closed walk is related to
the product of all links constituting that walk as
\be
l_1 \cdots l_N = q^{A(l_1, \ldots, l_N)} \;.
\label{area}
\ee

This can be simply understood by defining a normally ordered
walk as
\be
\{
y^{-1},y^{-1},\ldots,y^{-1},y,y,\ldots,y,x,x,\ldots,x,x^{-1},x^{-1},\ldots,x^{-1}\}\:. \label{normord}
\ee
Obviously, the area enclosed by such a walk vanishes,
and the product of its links is 1.
Calculating the area enclosed by an arbitrary walk, then,
reduces to normally ordering the product of its links and summing up all
the powers of $q$ generated by noncommuting links
being ``carried through'' each other.
E.g., a $1\times1$ square encircled anticlockwise
is $y^{-1}xyx^{-1} = qy^{-1}yxx^{-1} = q$,
and the  area is 1.

A closed walk of length $N$ containing $M$ instances of $x$
contains the same number of $x^{-1}$'s and,
correspondingly, $\frac{N}{2} - M$ of each of $y$'s and $y^{-1}$'s.
The number of such closed walks is
\be\label{no}
C_{N,M} = \frac{N!}{M!^2 (\frac{N}{2}-M)!^2}
\ee
($N!$ permutations of all the links, divided by permutations of
identical links), hence the total number of closed walks of length $N$ is
\be
C_N = \sum_{M=0}^{N/2} C_{N,M}={N\choose N/2}^2\:.
\label{P_N}
\ee
We are interested in finding the number $C_N(A)$ of closed walks
of length $N$ enclosing an algebraic area $A$; the area probability
distribution is then given by
\be\label{finebis}
P_N(A) = \frac{C_N(A)}{C_N} \;.
\ee
We will search for the generating function of said distribution,
\be\label{fine}
Z_N(q) = \sum_{A=-\infty}^\infty C_N(A)\, q^A \;.
\ee

To elucidate the method, consider at first the set of walks obtained from
(\ref{normord}) by shuffling only the $x$'s and $y$'s while leaving
the $x^{-1}$'s and $y^{-1}$'s in place.
For a given $M$ as defined above, the part of the walk formed by
the $x$'s and $y$'s is a ``staircase walk'' of width $M$ and height
\be
L = \frac{N}{2}-M \; ;
\ee
for brevity, call it an $(M,L)$ staircase walk.
Now invoke the $q$-binomial theorem \cite{Andrews}:
If $x$ and $y$ satisfy Eq.~(\ref{qcomm}), then 
\be
(x + y)^{N/2} = \sum_{\stackrel{\scriptstyle M,L}{M+L = N/2}} Z_{M,L}(q) y^L x^M \;,
\label{qbin}
\ee
with $Z_{M,L}(q)$ being the $q$-binomial coefficient:
\be Z_{M,L}(q) = 
{M + L \choose L}_q  \equiv \frac{[M+L]_q!}{[M]_q! [L]_q!} \;,
\ee
where
\be
[L]_q!  =  \prod_{i=1}^{L}{1-q^{i}\over 1-q}
 =  1 (1+q) (1+q+q^2) \cdots (1 + q + \ldots + q^{L-1}) \;.
\ee
The function $Z_{M,L}(q)$ is the generating function of the
area distribution of $(M,L)$ staircase walks (more precisely,
the closures of those walks, obtained by appending $x^{-M}$
to the end of a walk and prepending $y^{-L}$ at the beginning), in the sense of Eq.~(\ref{fine}).
Indeed, the LHS of (\ref{qbin}) is the sum of all combinations
of products of $x$'s and $y$'s, with any ordering, such that the
total number of $x$'s and $y$'s is $N/2$.
The number of $(M,L)$ walks within this set that yield a
multiplier $q^A$ when normally ordered, i.e., turned into $y^L x^M$,
is the coefficient at $q^A$ in $Z_{M,L}(q)$.
Respectively, the generating function of the area distribution
of all staircase walks of length $N/2$ is $\sum_{M+L = N/2} Z_{M,L}(q)$.

Consider now rewriting $(x + y)^{N/2}$ as $ (x + y)(x + y)^{N/2-1}$ and expand the binomials on both sides using Eq.~(\ref{qbin}) to find
\be
Z_{M,L}(q) = Z_{M,L-1}(q) + q^LZ_{M-1,L}(q) \;,
\label{recsimple}
\ee
with the initial condition $Z_{0,0}(q) = 1$.

There is  a simple physical interpretation of $Z_{M,L}(q)$.
Constructing a walk consisting of $M$ instances of $x$ and
$L$ instances of $y$ amounts to
distributing the $M$ $x$'s among the $L+1$ slots between the $y$'s
(slot number $0$ being to the right of the last $y$,
slot number $k = 1, \ldots, L$ to the left of the $k$-th $y$,
counting the latter from right to left).
Denote by $m_k$ the number of $x$'s put into slot number $k$;
there is a constraint $\sum_{k=0}^L m_k = M$.
To normally order the walk, i.e., to move all the $x$'s to the right,
one has to carry each of the $m_k$ $x$'s through $k$ $y$'s,
hence the total area is $\sum_{k=0}^L k m_k$.
Now think of the $k$-th slot as a single-particle state with energy $\varepsilon_k=k$
and of the $x$'s as bosons distributed among those states
with occupation numbers $m_k$.
Then the number of walks with an algebraic area $A$ is 
the multiplicity of the multiparticle level with energy $A$.
Respectively, the generating function of the area distribution, $Z_{M,L}(q)$,
is equal to the partition function of $M$ bosons
in $L+1$ single-particle states with energies $0, \ldots, L$,
with $q = \rme^{-1/T}$.
The recurrence relation (\ref{recsimple}) can be interpreted as
the first addend on the RHS being the sum over all multiparticle states
in which the highest single-particle level (whose energy is $L$)
is empty; the second addend being
the sum over all states in which that level is occupied by at least one boson.

Note that levels and bosons can actually be interchanged,
since $Z_{M,L}(q) = Z_{L,M}(q)$.
At $q=1$ (infinite temperature) relation (\ref{recsimple})
becomes the Pascal triangle equation, and
$Z_{M,L}(1) = {M + L \choose L}$ is  the total
number of multiparticle states, or walks.

Now consider all possible closed  walks of length $N$.
In order to obtain the generating function of their area distribution,
one can generalize  (\ref{qbin}) as
\be
(x + y + x^{-1} + y^{-1})^N =
\sum_{\stackrel{\scriptstyle M_1,M_2,L_1,L_2}{M_1+M_2+L_1+L_2 = N}}
Z_{M_1,M_2,L_1,L_2}(q) y^{-L_1}y^{L_2} x^{M_1} x^{-M_2}  \;.
\label{qbingen}
\ee
Proceeding as above, one concludes that
$Z_{M,M,\frac{N}{2}-M,\frac{N}{2}-M}(q)$ is the generating
function of the area distribution of closed walks containing
$M$ instances of $x$ (and as many $x^{-1}$) and $N/2-M$ instances of $y$
(and as many $y^{-1}$).
For all closed walks of length $N$, one has to sum
over all possible values of $M$:
\be
Z_N(q) = \sum_{M=0}^{N/2} Z_{M,M,\frac{N}{2}-M,\frac{N}{2}-M}(q) \;.
\label{ZN}
\ee
The recurrence relation for $Z_{M_1,M_2,L_1,L_2}(q)$  generalizes  Eq.~(\ref{recsimple}):
\bea
Z_{M_1,M_2,L_1,L_2}(q) & = &  Z_{M_1,M_2,L_1-1,L_2}(q)+ Z_{M_1,M_2,L_1,L_2-1}(q) \nonumber \\
&& {} + q^{L_2-L_1}Z_{M_1-1,M_2,L_1,L_2}(q) + q^{L_1-L_2}Z_{M_1,M_2-1,L_1,L_2}(q) \;,
\label{recbis}
\eea
with the initial condition $Z_{0,0,0,0}(q) = 1$.

The problem of finding the
area distribution has been  reduced to solving this recurrence relation.

\section{Results}

\sloppypar
Obvious symmetry considerations imply
$Z_{M_1,M_2,L_1,L_2}(q)=Z_{L_1,L_2,M_1,M_2}(q)$, as well as
$Z_{M_1,M_2,L_1,L_2}(q)=Z_{M_2,M_1,L_1,L_2}(1/q)$ (mirror reflection).
In general,
\be Z_{M_1,M_2,L_1,L_2}(q)=\sum_{A=-A_-}^{A_+}C_{M_1,M_2,L_1,L_2}(A)q^A\;,
\ee 
where the $C$'s  are integers and
\be 
A_{\pm} = \max (M_1,M_2)\max (L_1,L_2)-\frac{|(M_{1}-M_2)(L_{1}-L_2)|\pm (M_{1}-M_2)(L_{1}-L_2)}{2} \;.
\ee
$Z_{M_1,M_2,L_1,L_2}(1)$ is  the 
number of walks involved, i.e., the multinomial coefficient
\be Z_{M_1,M_2,L_1,L_2}(1)={ (M_1+M_2+L_1+L_2)!\over M_1!M_2!L_1!L_2!}\ee 
[Eq.~(\ref{no}) corresponds to the case $M_1=M_2=M$ and $ L_1=L_2=N/2-M$].
At first order in $q-1$,
\be
Z_{M_1,M_2,L_1,L_2}(q) = Z_{M_1,M_2,L_1,L_2}(1)
\left[1-{(M_{1}-M_2)(L_{1}-L_2)\over 2}(q-1) + \ldots \right]  .
\ee

Further, when one of the subscripts vanishes,
for example $M_1=0$, one has\footnote{When $q=1$, the identity 
\be \sum_{k=0}^{\min(L_1,L_2)} \left[ {M_2 + L_1 + L_2 \choose k} - {M_2 + L_1 + L_2 \choose k-1} \right]{M_2 + L_1 -k \choose M_2}{M_2 +  L_2-k \choose M_2}={ (M_2+L_1+L_2)!\over M_2!L_1!L_2!}\nonumber\ee
follows.}
\be\label{solut}
Z_{0,M_2,L_1,L_2}(q) = \sum_{k=0}^{\min(L_1,L_2)} \left[ {M_2 + L_1 + L_2 \choose k} - {M_2 + L_1 + L_2 \choose k-1} \right]
Z_{M_2,L_1-k}\big(q\big) Z_{M_2,L_2-k}(\frac{1}{q}) \;.
\ee
It can be  verified by direct calculation, using Eq.~(\ref{recsimple}), that
$ Z_{0,M_2,L_1,L_2}(q)  =   Z_{0,M_2,L_1-1,L_2}(q)+ Z_{0,M_2,L_1,L_2-1}(q)
  + q^{L_1-L_2}Z_{0,M_2-1,L_1,L_2}(q)$.
Furthermore 
\be
(x^{-1} + y + y^{-1})^N =
\sum_{\stackrel{\scriptstyle M_2,L_1,L_2}{M_2+L_1+L_2 = N}}
Z_{0,M_2,L_1,L_2}(q) y^{-L_1}  y^{L_2} x^{-M_2}\;
\label{qbinter}
\ee
takes place. This generalizes the $q$-binomial theorem onto the case of three addends.


Equation (\ref{solut}) relates to the subset of 
closed  walks  obtained from (\ref{normord})
by first moving the $x$'s to the end of the walk and then shuffling  the $x^{-1}$'s, $y$'s and $y^{-1}$'s. Following the same line of reasoning as for the staircase walks,
the slots between the $y$'s and $y^{-1}$'s correspond to $L_1+L_2+1$ single-particle states,
again numbered from 0 onwards, but now their energies satisfy
$\varepsilon_k=\varepsilon_{k-1}\pm 1$,
where the sign coincides with the sign of the power of $y$ at the $k$-th position from right
(the initial condition is still $\varepsilon_0=0$).
Hence, the energy of the last state is $\varepsilon_{L_1+L_2} = L_1-L_2$.
The $M_2$ instances of $x^{-1}$ have to be distributed in all possible single-particle spectra stemming from all possible arrangements of the $y$'s and $y^{-1}$'s,
and Eq.~(\ref{solut}) is the partition function
of $M_2$ bosons in such a single-particle spectrum,
summed over all possible single-particle spectra.

In the special case  $L_1=L_2=L$, the single-particle spectra are in one-to-one correspondence with all possible bilateral Dyck paths of length $2L$. 
The partition function in question is then related to the statistics of  such paths.




In the absence  of a general closed-form solution 
for $Z_{M_1,M_2,L_1,L_2}(q)$,  Eq.~(\ref{recbis}) has to be solved iteratively (see Table 1).

\begin{center}
\begin{tabular}{|r|c|c|c|c|c|c|}
\hline
& $N = 2$ & 4 & 6 & 8 & 10 & 12 \\
\hline
$A = 0$ & 4 & 28 & 232 & 2156 & 21944 & 240280 \\
\hline
$\pm 1$ & & 4 & 72 & 1008 & 13160 & 168780 \\
\hline
$\pm 2$ & & & 12 & 308 & 5540 & 87192 \\
\hline
$\pm 3$ & & & & 48 & 1560 & 33628 \\
\hline
$\pm 4$ & & & & 8 & 420 & 11964 \\
\hline
$\pm 5$ & & & & & 80 & 3636 \\
\hline
$\pm 6$ & & & & & 20 & 1200 \\
\hline
$\pm 7$ & & & & & & 264 \\
\hline
$\pm 8$ & & & & & & 72 \\
\hline
$\pm 9$ & & & & & & 12 \\
\hline
\end{tabular}
\end{center}
\begin{center}
Table 1:  Nonzero values of $C_N(A)$
for $N \le 12$.
\end{center}

Note that:

(i) if $\frac{N}{2}$ is even,  the maximal possible area
is $\frac{N^2}{16}$,
and $C_N(\frac{N^2}{16}) = N$.  The maximal  area is obtained
for an anticlockwise square walk $(\frac{N}{4} \times \frac{N}{4})$,
and all such   walks are the $N$ cyclic
permutations of 
$\{x,\ldots,x,y,\ldots,y,x^{-1},\ldots,x^{-1},y^{-1},\ldots,y^{-1}\}$;

(ii) if $\frac{N}{2}$ is odd,  the maximal possible area
is $\frac{N^2-4}{16}$,
and $C_N(\frac{N^2-4}{16}) = 2N$.
The maximal area is obtained for anticlockwise rectangular walks $(\frac{N-2}{4} \times \frac{N+2}{4})$
and $(\frac{N+2}{4} \times \frac{N-2}{4})$, and all such walks are the $N$
cyclic permutations for one configuration and the $N$
cyclic permutations for the other one.

(iii) knowing these values,  one can calculate the $k$-th moment of the area distribution as a function of $N$
\be
R_k(N) = \sum_{A=-\infty}^{\infty}P_N(A) A^k\;.
\ee
The results for $R_2(N)$ and $R_4(N)$
coincide with those of  Ref.~\cite{Mingo}.

\section{Asymptotic limit}

To make the connection  with the Harper-Hofstadter model,
define a lattice site 
$m= (m_1, m_2)$, and magnetic translation operators $W(m)$, which satisfy
\be W(m)W(m')=W(m+m')\rme^{\rmi\gamma (m_1 m'_2-m_2m'_1)}\label{mag}\;,\ee
where $\gamma=2\pi\phi/\phi_0$ with $\phi$ the flux
of the magnetic field per unit cell,
$\phi_0$ the flux quantum.
Equation~(\ref{mag}) is  the Harper-Hofstadter    counterpart
of Eq.~(\ref{qcomm}) for the algebraic area distribution of random walks.

The Harper-Hofstadter Hamiltonian is \cite{Bell, Hofstadter}
\be H=\sum_{|m_1|+|m_2|=1}W(m)\;,\ee
such that
\be {\rm Tr}\,H^N=\sum\rme^{\rmi\gamma A}\;,\ee
where the trace on the LHS is per unit cell
and the summation on the RHS is over all closed walks of length $N$.
Setting  $q\to \rme^{\rmi x/N}$, the mapping of the problem of random walks
onto the Harper-Hofstadter problem follows as
\be\label{map} Z_N(\rme^{\rmi x/N})={\rm Tr}\,{H^N}|_{\gamma=x/N} \;.\ee
From Eqs.~(\ref{finebis})--(\ref{fine}) one finds  that
\be {Z_N(\rme^{\rmi {x/ N}} )\over Z_N(1)}=
\sum_{A=-\infty}^\infty P_N(A)\, \rme^{\rmi A{x/ N}}\;, \ee
thus establishing, via (\ref{map}), a one-to-one correspondence
between the algebraic area probability distribution $P_N(A)$
and the $N$-th moment of the Harper-Hofstadter spectrum ${\rm Tr}\,{H^N}$.

Evaluating ${\rm Tr}\,{H^N}|_{\gamma=x/N}$
in the large $N$ limit \cite{Bell}, one concludes that
\be
Z_N( \rme^{\rmi x /N} ) = {4^{N+1}\over 2\pi N}\frac{x/4}{\sinh(x/4)}
\left[ 1 - \frac{1}{2N}\,\frac{(x/4)^2}{\sinh^2(x/4)} + O(1/N^2)\right]
\label{mapping}\ee
must take place.
Obviously, for a finite $N$, (\ref{mapping}) cannot hold
for all $x$, since
the LHS is periodic with period $2\pi N$.
Still, for $x < \pi N$, calculating  $Z_N$ from
Eqs.~(\ref{ZN})--(\ref{recbis}) results in an excellent agreement,  even
for rather small values of $N$ (see Fig.~1).

\begin{figure}
\begin{center}
\includegraphics[scale=1]{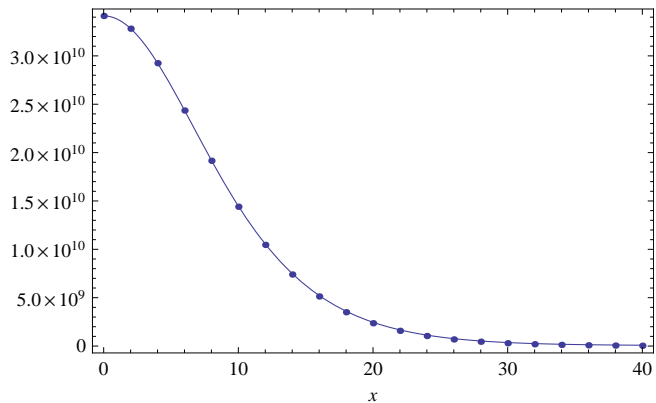}\label{fig1}
\caption{ $Z_N( \rme^{\rmi x /N} )$ (line) and
${\rm Tr}\,{H^N}|_{\gamma=x/N}$, the RHS of Eq.~(\ref{mapping}) (dots),
for $N=20$.}
\end{center}
\end{figure}

\section{Conclusion}

The area distribution of closed random walks on a square lattice
stems from the noncommutative nature of the links that form such walks.
This is reflected in the fact that the generating function
of that distribution is directly connected with
the moments of the spectrum of the Harper-Hofstadter Hamiltonian ---
a sum of four nearest-neighbor magnetic translation operators,
which become noncommutative in the presence
of an external magnetic field.
Clearly, the recurrence relation (\ref{recbis})
for the generating function can be interpreted in terms of  multi-body partition functions, as it has been done for staircase walks. 
The complexity of the 
 Harper-Hofstadter spectrum is encoded in some way in this recurrence relation.

A generalization of the $q$-binomial theorem has been obtained in the case of three addends. In the case of four addends, closed-form expressions for the generating
function might involve $q$-deformed multinomial coefficients ---
as suggested by the three-addend solution.

{Acknowledgements: S.M. would like to thank the LPTMS in Orsay for the hospitality during the completion of this work. S.O. would like to thank A. Comtet for interesting discussions.}

\end{document}